\documentclass[aps,tightenlines,prc,nofootinbib]{revtex4}
\usepackage{epsfig,graphics,color}
\usepackage{graphicx}
\usepackage{dcolumn}
\usepackage{bm}
\usepackage{amsmath,amssymb}
\usepackage{eucal}

\newcommand \tie {{\it i.e.}}

\newcommand \f {\not\!}

\newcommand \kd  {\delta}
\newcommand \ra  {\rightarrow}

\newcommand \h {\theta}

\newcommand \vk {\vec{k}}
\newcommand \vl {\vec{l}}
\newcommand \vq {\vec{q}}

\newcommand \vp {\vec{p}}

\newcommand \g {\gamma}

\newcommand \e {\epsilon}

\newcommand \p {{\prime}}

\newcommand \x {\cdot}
\newcommand \hf {\frac{1}{2}}
\newcommand \A {\alpha}

\newcommand \lc {\langle}
\newcommand \rc {\rangle}
\newcommand \prt {\partial}

\newcommand \nt {\noindent}

\newcommand \ld {\lambda}

\newcommand \ml {\mathcal{L}}
\newcommand \gmn {g^{\mu \nu}}

\newcommand \bvec{\left( \begin{array}{c} }
\newcommand \evec{\end{array} \right)}
\newcommand \tr {\mbox{{\bf Tr}}}
  
\newcommand \bea{\begin{eqnarray} }
\newcommand \eea{\end{eqnarray} }
\newcommand \nn {\nonumber}
\newcommand \be {\begin{equation}}
\newcommand \ee {\end{equation}}
\newcommand \epem {$e^+ e^-$}
\newcommand \mbx {\mbox{}}

\newcommand \psibar {\bar{\psi}}
\newcommand \ata {& \times &}

\newcommand \slm {\sum\limits}
\newcommand \pbs {\not\!p_\perp}
\newcommand \dps {\not\!D_\perp}

\newcommand \pb {\vec{p}_\perp}
\newcommand \xb {\vec{x}_\perp}
\newcommand \kb {{\vec k}_{\perp}}

\newcommand \ks {\not\!k_\perp}
\begin{document}

\title{Extending Soft-Collinear-Effective-Theory to describe hard jets in dense QCD media}

\author{Ahmad Idilbi}
\affiliation{Department of Physics, Duke University, Durham, NC 27708, USA}

\author{Abhijit Majumder\footnote{Current Address: Department of Physics, 191 W. Woodruff Ave., The Ohio State University, Columbus OH 43210.} }
\affiliation{Department of Physics, Duke University, Durham, NC 27708, USA}

\date{ \today}

\begin{abstract}
An extension to the Soft-Collinear-Effective Theory (SCET) description of hard jets 
is motivated to include the leading contributions between the propagating partons within the 
jet with partons radiated from a dense extended medium. The resulting effective Lagrangian, containing  
both a leading and a power suppressed (in the hard scale $Q^2$) contribution, arises primarily 
from interactions between the hard collinear modes in the jet with Glauber modes from the medium.  
In this first attempt, the interactions between the hard jet and soft and collinear partonic modes have 
been ignored, in an effort to focus solely on the interactions with the Glauber modes. 
While the effect of such modes on vacuum cross sections are suppressed by powers of the hard scale 
compared to the terms from the SCET Lagrangian, such sub-leading contributions are enhanced by the extent 
of the medium and result in measurable corrections. The veracity of the derived 
Lagrangian is checked by direct comparison with known results from full QCD 
calculations of two physical observables: the transverse momentum broadening 
of hard jets in dense media and a reanalysis of the transverse momentum 
dependent parton distribution function (TMDPDF). 
\end{abstract}


\maketitle


 \section{introduction}


The study of hard jets in QCD is now a considerably mature science. Experiments 
at \epem annihilation, Deep-Inelastic Scattering (DIS) and $p$-$p$ machines have yielded 
a wide array of measurements on a variety of jet observables including single particle 
production, multi-particle  correlations as well as event shapes. On the theory 
side, sophisticated factorization theorems have been written down which factorize 
the final state ``jet function'' from the initial state and the hard cross section 
at leading twist~\cite{ster}. 
In the derivation leading to such factorization 
theorems~\cite{Ellis:1978sf,Libby:1978qf,Collins:1981uw,Collins:1983ju,Collins:1985ue} 
the infinite class of Feynman diagrams are subjected to a Landau analysis. 
Regions of momentum space which yield pinch singularities are identified.  
These represent the leading contributions to such processes and may be 
decomposed into classes of Feynman diagrams which in turn allow for proofs of 
factorization. An alternate and equivalent 
approach has recently been afforded by the methods of effective field theories  
such as the Soft-Collinear-Effective-Theory~(SCET)~\cite{Bauer:2000ew,Bauer:2000yr,Bauer:2001ct,Beneke:2002ph,Chay:2002vy}. 
While not specifically devised to 
re-derive factorization, SCET presents a formalism where the analysis resulting in the 
identification of the leading contributions may be carried out within the QCD Lagrangian resulting 
in the derivation of an effective Lagrangian which is only applicable to processes within the 
prescribed kinematic regime.
In such a formalism, factorization occurs at the level of the Lagrangian and at the level of operators~\cite{Bauer:2001yt,Bauer:2002nz}. 
The Feynman rules which 
arise from an expansion in a small parameter $\lambda$ may then be used to 
systematically study hard processes. 

Power corrections to hard processes in vacuum are suppressed in the presence of a hard 
scale $Q^2 \gg \Lambda^2_{QCD}$. However, there exist scenarios where a specific 
set of power corrections (often arising from operators with higher twist) may be enhanced 
and become non-negligible compared to the leading process. One example of this is the 
case of single spin asymmetry in DIS, where leading twist processes yield vanishing results~\cite{Ji:2006br}.
Another example, where the leading twist term does not vanish but power corrections may become enhanced 
is in the presence of a medium
is in the case of DIS on 
a large nucleus~\cite{Qiu:1990xx,Qiu:1990xy}. The inclusive cross section receives
contributions from power suppressed operators which are enhanced 
by a factor $A^{1/3}$ arising from the length of a large nucleus with mass number $A$~\cite{Luo:1992fz,Luo:1994np}.

In semi-inclusive processes such as single hadron inclusive events in DIS on a large nucleus, a hard jet 
is formed in the collision of the virtual photon with a hard quark. This jet then begins to shower and 
lose virtuality on its way to hadronization. Some part of this space-time evolution occurs within the 
nuclear medium. Multiple scattering of the jet in the medium modifies the final distribution of high momentum 
hadrons emanating from such a 
hard jet~\cite{Guo:2000nz,maj04e}. Experimental measurements of single and multi-hadron production from such modified jets 
and their comparison with jets produced in DIS on a proton or in $p-p$ collisions allow one to quantify the 
gluon distribution in dense extended QCD media~\cite{Airapetian:2000ks}.

This modification depends on a class of higher twist 
operators evaluated in the nuclear medium. While there exists considerable information regarding 
the ground state (nucleon) structure of large nuclei which may be invoked in the modeling of these higher 
twist operators, there exists practically no such information regarding the bulk structure of the 
deconfined matter produced in high-energy heavy-ion collisions. 
The modification of hard jets, in the deconfined matter produced in heavy-ion collisions, has 
assumed center stage in the experimental program as the primary probe of the structure of the 
produced matter~\cite{white_papers}. 
This is due in part to  
the dramatically large effects seen in comparison with cold confined matter as well as the 
possibility for a first principles computation of this modification from perturbative QCD (pQCD). 
There are many approaches to this calculation, all involving a different set of approximations 
about the medium~\cite{Wang:2002ri,Majumder:2007ae,Baier:1996kr,Gyulassy:2000fs,Wiedemann:2000za}.
For a review of the different approaches see Ref.~\cite{Majumder:2007iu}.

While the benefits of an effective field theory description of power corrections to 
hard processes in QCD cannot be overstated, to date there has not been a single attempt to incorporate 
the leading effects of medium enhanced higher twist within an effective theory formalism such as SCET.
There exists an effective theory description of the dense deconfined matter in the limit of very high 
temperature. Here, a consistent effective theory of the medium without a jet is set up first, then the hard 
jet is assumed to have interactions with the soft field in the medium which are similar to  
those encountered by a hard thermal parton~\cite{Arnold:2002ja}.
This article will take a different route, by trying to extend an existing effective theory of jets in vacuum to incorporate the 
effects of scattering in a medium. In what follows, we undertake the 
simplest extension to an SCET like formalism in the presence of an extended QCD medium.

In Sec.~II, the emergent scales in the problem will be discussed and the presence of a 
new mode, called the Glauber mode, not present in the current vacuum implementation of SCET will be motivated. 
In Sec.~III, an effective Lagrangian which includes the interaction of such Glauber modes with collinear quarks will be 
derived from the QCD Lagrangian. In this first attempt, we will ignore the further interactions between these 
Glauber modes and the soft and collinear gluon modes of the usual SCET Lagrangian. A new set of Feynman 
rules arising from such an effective Lagrangian will be outlined and their equivalence with the Feynman rules of 
full QCD demonstrated at an amplitude by amplitude level. In Sec.~IV and V, the Feynman rules will be used to 
compute cross sections in physical processes; two examples will be dealt with: the transverse 
broadening of jets in DIS on large nuclei and the  transverse-momentum-dependent-parton-distribution-function (TMDPDF). It will be shown explicitly that, in light-cone gauge, the Glauber gluons give rise to the transverse gauge link that enters the definition of the fully gauge invariant TMDPDF.
The results obtained will be compared with published 
calculations in full QCD. Concluding discussions will be presented in Sec.~VI.


 \section{The energy scales from a dense medium }

\begin{figure}[htbp]
\resizebox{3in}{2in}{\includegraphics[0in,0in][6in,4in]{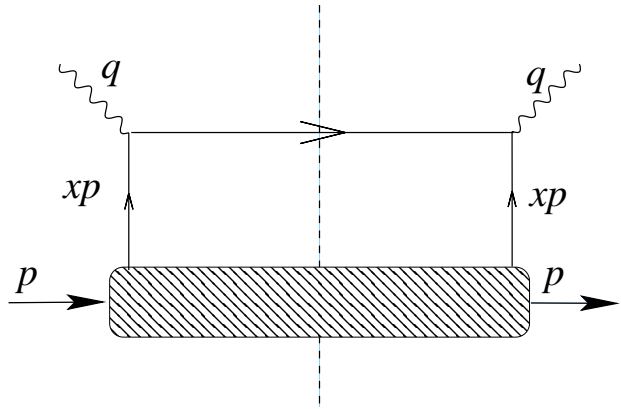}} 
    \caption{ Single-inclusive DIS on a nucleon or a large nucleus.}
    \label{fig1}
\end{figure}

Consider the deep inelastic scattering (DIS) of a hard photon with momentum $q$ and virtuality $q^2 = -Q^2$ 
on a nucleon with momentum $p$ 
in vacuum or contained within a large nucleus. 
In the Breit frame, the photon has the momentum components
\bea
q \equiv [q^+, q^-,\vq_\perp] = [Q^2/2q^-, q^-,0 ] \sim Q(-1,1,0). \label{photon_scaling}
\eea 
The off-shellness of the photon $Q$ is taken as a representative of the hard scale in the process. This strikes a hard, 
almost on-shell quark with a large momentum in the $+z$ direction or a large $(+)$-component of momentum,
\bea
p_i \equiv (x_B p^+, p_i^- , \vp_{i,\perp}) \sim Q(1,\lambda^2,\lambda). \label{in_jet_scaling}
\eea
The number of partons in the infinite momentum frame which carry an $x_B \sim 1$ fraction 
of the nucleon momentum is rather small and $\lambda$ is a small dimensionless variable ($\lambda \ra 0$). 
We then trigger on events where an 
almost on-shell jet is produced in the final state (see Fig.~\ref{fig1}). This partonic jet moves with large 
momentum in the $(-)$-direction,
\bea
p_f \equiv (p_f^+, q^-, \vp_{f,\perp}) \sim Q(\lambda^2 , 1, \lambda),  \label{out_jet_scaling}
\eea
where, at leading order, $\vp_{f,\perp} = \vp_{i,\perp}$.

In the case of DIS on a single nucleon, this partonic jet immediately escapes the medium 
and eventually after a time $\sim 1/(\lambda^2Q)$~\footnote{The formation time of the radiation may be estimated from the virtuality of the jet $\sim \lambda Q$ and the boost of the jet $\sim \lambda^{-1}$.} would have decayed into 
multiple partons of lower invariant mass and eventually turns into a jet of hadrons. The 
invariant mass of the final jet is $m \sim \lambda Q$
and its total forward energy from all produced hadrons which have arisen from this jet is 
\bea
E = q^-/\sqrt{2} \sim Q.
\eea
In this article, we initiate  the construction of the effective theory for the propagation of such jets through the dense 
matter within nucleons and large nuclei.

The first step in such an endeavor is the assignment of relations between 
all relevant dimensionful quantities that 
appear in the problem. We assume that the scaling variable $\lambda$ is so chosen 
that perturbation theory may be applied down to momentum transfer scales at or above $\lambda^2 Q$. 
The medium, also introduces its own set of scales, e.g., the mass of a nucleon $M_N \simeq 1$GeV. This is assumed to 
scale with a new scaling variable $\mu$ such that it is 
comparable to $\Lambda_{QCD}$ and in general much smaller than the 
soft perturbative scale, \tie,  
\bea
M_N \sim \Lambda_{QCD} \sim \mu Q \lesssim \lambda^2 Q.
\eea 
One may immediately surmise that in the 
Breit frame when $x_B \sim 1$, and $p^+ \sim Q$, the boost or $\g$-factor is of order $\mu^{-1}$.
While the inverse size of the nucleon may be thought of as an even softer scale, in this effort, we will 
assume that the hard scales $Q, \lambda Q $ are much harder than the medium scale $\mu Q$ and thus 
the inverse length will be assumed to be of the order of the mass of the nucleon,
\bea
l_N \sim \frac{1}{\mu Q}, 
\eea
All our considerations will be carried out in the Breit frame. The scaling introduced at the beginning of this 
section regarding the momentum components of the incoming and outgoing partons were set up in the 
Breit frame. Thus the nucleon (or medium) will have to be boosted to this frame.
The ensuing boost to the Breit frame will lead to the contracted length of the nucleon, 
\bea
\frac{l_N }{ \g} \sim \frac{1}{ Q}.  
\eea
The very introduction of an alternative soft scale such as $\mu Q$ may lead the reader to imagine a 
much more complicated effective theory which will manifestly involve both $\lambda$ and $\mu$ 
and will require a relation between these two scaling variables. However, as will be demonstrated in
the next two sections, with specific examples, it is possible to construct an effective theory in dense matter 
involving only the hard scale $Q$ and the vacuum scaling variable $\lambda$. 

The construction of an in-medium effective theory which depends on only the hard scale and the 
scaling variables from the vacuum theory has one further requirement. In specific examples, such as 
in Sec.~IV, certain in-medium matrix elements will be enhanced by media with sizes much larger than 
a nucleon, e.g., in the case of large nuclei or a deconfined quark gluon plasma (QGP). 
In the case of large nuclei (with mass number $A \gg 1$), the enhancement factor is usually the nuclear 
length in units of the nucleon length i.e., $A^{1/3}$. These situations 
will require that the enhancement be expressed in powers of $\lambda$, i.e., $A^{1/3} \sim \lambda^{-n}$. 
The number $n$ is so far unspecified and will turn out to be observable dependent.

The particular choice of scaling of nucleon size and momentum lead to certain obvious physical consequences: 
In the Breit frame, valence (large $x$) partons carrying order one fractions of the forward momentum of the nucleon, 
have momenta that may be expressed as (by simply boosting momenta of the order of $M_N$),
\bea
k \sim Q (1,\lambda^2, \lambda ) ,
\eea
These partons  
will be found to be completely confined within nucleons. The off-shellness of these partons $\sim \lambda^2 Q$
is very small and, as a result, for processes which involve momentum transfers of the order of $\lambda Q$ or larger, 
these radiated partons may be considered as asymptotic in-states. 
In the following, we will often ignore discussion of the $(-)$-components 
of the partonic momenta in the in-state; these play almost no role in the computation of jet modification and 
the transverse momentum dependent structure functions.

\begin{figure}[htbp]
\begin{center}
\hspace{0cm}
\resizebox{1.5in}{1.5in}{\includegraphics[0in,0in][3in,3in]{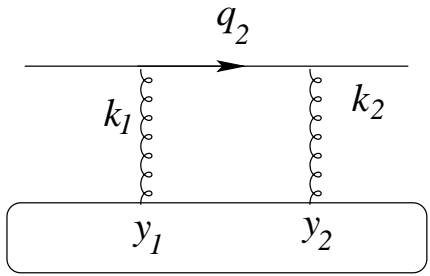}} 
\vspace{0.25cm}
\caption{ The multiple scattering of a produced jet in Deep-Inelastic scattering. }
   \label{fig2}
 \end{center}
\end{figure}

Due to interactions, a variety of gluons may be radiated from these valence partons with momenta 
constrained by overall energy-momentum conservation and determined by the kinematics of the 
process being triggered on. In the case of transverse broadening, the type of radiated gluon which 
plays a leading role will be those with momenta which scale as 
\bea
k \sim Q (\lambda^2,\lambda^2 , \lambda ). \label{glauber_scaling}
\eea
Gluons with momenta which scale as in the above equation are referred to as Glauber gluons, or gluons in the 
Glauber region. 
The role of Glauber gluons in transverse broadening (as well as in transverse momentum dependent structure functions) 
may be easily understood in the case of DIS with a hard jet in the final state. The momenta of the produced quark jet 
scales as in Eq.~\eqref{out_jet_scaling}. Imagine the multiple scattering of the struck quark off the remnants of the nucleon or
nucleus. The diagrams under consideration are of the form of Fig.~\ref{fig2}. 
In order for the produced jet to escape from the nucleon or nucleus without undergoing any induced radiation, 
the interactions with the nucleon have
to be such that they do not induce a major change in the off-shellness of the quark. In order to see how this comes about we explicitly 
write out the expression for the $q_2$ propagator where $q_2 = q + p_i + k_1$,  represents the 
quark propagator, which, after the hard scattering 
(with momentum $q + p_i$)  scatters off one extra gluon with 
momentum $k_1$, (in order to simplify the expression we assume that 
$p^i_\perp = 0$ and $p_i^+ = x_Bp^+ = - q^+$) 
in Fig.~\ref{fig2}:
\bea 
S(q_2) &=& \frac{ \g^+ q^- +  \g^\perp \vk^1_\perp + \g^- k_1^+  }{ 2q^- ( k_1^+ )  - | \vk^1_\perp |^2 } \nn \\
&\simeq& \frac{ \g^+ q^-}{2q^- ( k_1^+ )  - | \vk^1_\perp |^2}.
\eea
Since, $q^- \sim Q$ and $k^1_\perp \sim \lambda Q$, the forward momentum has to scale as $k_1^+ \sim \lambda^2 Q$ for 
the jet to remain off-shell by no more than $\lambda^2 Q^2$. If the forward momentum scales with a higher power, 
\tie,  $k^+ \sim \lambda Q$, 
this will cause the jet to go off-shell by $\lambda Q^2$ and lead to the radiation of momenta with large transverse momenta 
$l_\perp \sim \lambda^{\frac{1}{2}} Q $. This process will lead to the radiative energy loss of the propagating quark and will 
be discussed in a future effort. The reader will note that the absorption of gluons which are collinear to the outgoing quark, i.e., with 
a momentum that scales as $Q(\lambda^2, 1 , \lambda)$ also do not raise the off-shellness of the quark beyond $\lambda^2 Q^2$. 
However, the number of such gluons emitted from a medium moving with a large collinear momentum in the $(+)$-direction is vanishingly small. 
Hence the effect of such partons will be ignored.

In the current manuscript, the focus will remain on the production of a single jet with non-zero 
transverse momentum.  
Since the number of gluons with a forward momentum $p^+ \sim \lambda^2 Q$  far exceeds those with $p^+ \sim \lambda Q$ for the 
same transverse momentum, 
the jet will tend to encounter multiple interactions with gluons with a soft forward momentum. These will result in the transverse 
broadening of the hard jet. The neglect of gluons with a larger (+)-component of momentum suppresses radiation from the 
hard parton.

With the power counting of the different momentum components identified, there remains the issue of determining the 
power counting (in terms of $\lambda$) of the 4-vector potential $A_a^{\mu}$ in this regime of momenta. In the case of 
effective theories of QCD in a vacuum, the power counting of the $A_a^{\mu}$ field is determined by an estimation 
of the powers of $\lambda$ from the gluon propagator. 
In the case of Glauber gluons such a methodology will yield 
incorrect results. In the Glauber region of momenta, the gluon propagator obtained from the full QCD Lagrangian 
is never on-shell.With the transverse momenta being larger than the light-cone components, Glauber gluons are 
always space-like off-shell. As demonstrated above, an shell collinear parton may interact with a Glauber gluon 
and have its transverse momentum changed by order $\lambda Q$ while still remaining on-shell. 
The simplest 
extension of an effective theory containing only collinear modes (the jet) to a medium with collinear modes 
travelling in the opposite 
direction (the target) will contain the interactions of the collinear jet parton with Glauber gluons radiated 
off the partons in the target which move in the opposite direction. The effective action has the simple 
form, 
\bea
S &=& \int d^4 x \,\, \left[ \mathcal{L}_{SCET}  +  j^a_\mu (x) {A^a}^\mu_G (x) \right] . \label{tot_action}
\eea 
where, $A^{\mu}_G$ is the Glauber field radiated from the target and $j^a_\mu$ is the current of the 
collinear partons from the jet. The kinetic and interaction terms for the collinear fields which 
constitute $j^a_\mu$ are contained within $\ml_{SCET}$~\cite{Bauer:2000yr} along with terms for the soft fields. 
In this effort, $\ml_{SCET}$ will not contain the collinear or soft modes from the target. These will 
be integrated out and included in effective Glauber field $A^{\mu}_G$.

Note that there is no kinetic term for the Glauber gluon; thus, it does not 
obey a classical equation of motion. It admits no mode expansion and is not quantized as the SCET 
modes. However such exchanges are included in the full QCD Lagrangian and are prevalent in the 
interaction of collinear modes from jet with those from the target.  
The Glauber field $A^{\mu}_G$ represents the effective classical field of the target partons.
The power counting of the various components of the Glauber field may be obtained from a calculation of 
its production in full QCD.  

As our goal here is simply to estimate the power counting of the various components of the Glauber 
field, we will ignore subtleties associated with non-linear terms in an interacting non-Abelian theory. 
We will estimate the $\ld$ power of the various components in an Abelian theory.
In a classical Abelian theory, the gauge field $A^{\mu}$ is obtained from a solution of the inhomogeneous 
Maxwell's equation. 
This is given as, 
\bea
A^\mu (x) &=& A_0^\mu (x) + \int d^4 y \mathcal{D}^{\mu \nu} (x-y) J_\nu (y), \label{linear_resp}
\eea
where, $A_0^\mu (x)$ is a solution of the homogeneous Maxwell's equation. 
By restricting the current to be collinear to the target direction, and insisting that 
the incoming partons in the target remain close to on-shell, we restrict the field $A^\mu (x)$ to only its 
Glauber component $A_G^\mu (x)$. In this region of momenta, there exists no solution of the homogeneous 
Maxwell's equation i.e., ${A_0}_G^\mu (x) = 0$. As a result, the Glauber field is obtained from the second 
term on the right hand side of Eq.~\eqref{linear_resp}. We now evaluate, the power counting of the 
various components of the Glauber field in covariant and light cone gauge.

\subsection{Covariant Gauge}

At leading order in covariant gauge, the gauge propagator $\mathcal{D}^{\mu \nu}$ is given as, 
\bea
\mathcal{D}^{\mu \nu} (x -y ) &=& \int \frac{d^4 k}{ (2 \pi)^4} \frac{ - i \gmn e^{-i k \x ( x - y )}  }{ k^2 + i \e}. 
\label{covariant_gauge_gluon}
\eea
In Eq.~\eqref{linear_resp}, $J^{\nu} (y) = \psibar (y) \g^\nu \psi (y)$ is the current of partons in the target which generates the gauge field. The fermionic operator may be decomposed as,
\bea
\mbx\!\!\!\psi (y) \!\!&=&\!\! \int \frac{d p^+ d^2 p_\perp }{(2\pi)^3 \sqrt{p^+ + \frac{p_\perp^2}{2p^+}  }} 
\sum_s u^s(p) a_p^s e^{-ip\x y} + v^s (p) {b^s_p}^\dag e^{ip \x y}
\eea
The scaling of the fermionic operator depends on the range of momentum which are selected 
from the in-state by the annihilation operator. Note that this influences both the scaling of the 
annihilation operator $a_p$ as well as the bispinor $u(p)$. The power counting of the annihilation 
operator may be surmised from the standard anti-commutation relation, 
\bea
\{ a_p^r , {a_{p^\p}^s}^{\dag} \} = (2 \pi)^3 \kd^3 (\vp - \vp^\p) \kd^{rs}.
\eea
and the power counting of the bispinor from the normalization condition,
\bea
\sum_s u_p^s \bar{u}_p^s = \f p = \g^- p^+ + \g^+ p^- - \g_\perp \x p_\perp.
\eea 
Substituting the equation for the current in Eq.~\eqref{covariant_gauge_gluon}, and integrating out 
$y$, we obtain the expression for the ($+$)-component of the gauge field:
\bea
\mbx\!\!\!\!\!\!\!A^+ \!\!\!&\simeq&\! \!\!\!\int\!\! \frac{d^3 p d^3 q}{(2\pi)^6  \sqrt{p^+} \sqrt{q^+}}  
\frac{-i e^{-i(p-q)\x x} }{ (p-q)^2} a_q^\dag a_p 
\bar{u}(q) \g^+ u(q). \label{A_+_in_cov_gauge}
\eea
If the incoming and out going momenta $p$ and $q$ scale as collinear momenta in the ($+$)-direction, 
i.e., $p \sim Q(1,\lambda^2, \lambda)$, then we get, $\kd^3 ( \vp - \vp^\p ) \sim [ \lambda^2 Q^3 ]^{-1}$, as one 
of the momenta will involve the large ($+$)-component and the remaining are the small transverse 
components. Thus the annihilation (and creation) operator scales as $\lambda^{-1} Q^{-3/2}$. Also in the spin 
sum $\f p \sim Q$ and thus $u(p) \sim u(q) \sim Q^{1/2}$. The $\g^+$ 
projects out the large ($\sim Q$) components in $u$ and  $\bar{u}$ in the expression $\bar{u}(q) \g^+ u(p)$. 
We also institute the Glauber condition that 
$p^+ - q^+ \sim \ld^2 Q$, $p^- - q^- \sim \ld^2 Q$ and $p_\perp - q_\perp \sim \ld Q$.

Using these scaling relations we correctly find that the bispinor scales as 
$\lambda Q^{3/2}$. However, to obtain the correct scaling of the gauge field $A^+$ 
one needs to institute the approximation that $q^+ = p^+ + k^+$ where $k^+ \sim \lambda^2 Q$.
This condition is introduced by insisting that the $(+)$ momentum of the incoming and outgoing 
state, which control the scaling of $a_q^\dag$ and $a_p$, are separated by $k^+ \sim \ld^2 Q$. This is used to 
shift the $dq^+ \ra d k^+$ and as a result we obtain the scaling of the $A^+$ field as $\ld^2 Q$.
Following a similar derivation, with the replacement $\g^+ \ra \g^\perp (\g^-)$ we obtain the scaling of the 
transverse and ($-$)-component of the gauge field as $A^\perp \sim \ld^3 Q$ and $A^- \sim \ld^4 Q$.

\subsection{Lightcone Gauge}

The power counting of the gauge field is gauge dependent. In this last subsection we surmise the power counting, in light-cone 
gauge, for the Glauber field. The primary difference with Eq.~\eqref{A_+_in_cov_gauge} is the gauge field  
propagator. In the positive light cone gauge: $n\x A = n^- A^+ = A^+=0$, the only non-zero components are $A_\perp$ and $A^-$. 
Note that a Glauber field with transverse momentum $k_\perp \sim \ld Q$ can only be radiated from a 
collinear parton without changing the direction of the collinear parton. For $A_\perp$, the dominant contribution to the 
power counting equation arises not from the $\gmn$ term in the numerator of the propagator, but rather from the 
$(k^\mu n^\nu + k^\nu n^\mu)/k^+$ term, i.e., 
\bea
\mbx\!\!\!\!\!\!\!A_\perp \!\!\!&\simeq&\! \!\!\!\int\!\! \frac{d^3 p d^3 q}{(2\pi)^6  \sqrt{p^+} \sqrt{q^+}}  
\frac{ i \left(  \frac{ (p_\perp - q_\perp) n^- }{ p^+ - q^+ }   \right) e^{-i(p-q)\x x}     }{ (p-q)^2}   a_q^\dag a_p 
\bar{u}(q) \g^+ u(q). \label{A_perp_in_cov_gauge}
\eea
Comparing this with Eq.~\eqref{A_+_in_cov_gauge}, we obtain that $A_\perp \sim \ld Q$. 
Similarly we obtain $A^- \sim \ld^2 Q$.
As a result, in light cone gauge, the ($\perp$)-component of the gauge field is much more dominant than 
in covariant gauge and we expect this to change the power counting of various terms in the effective Lagrangian. 
Similar power counting arguments may also be surmised from the explicit expressions presented in Ref.~\cite{ji1} and references therein. See also the discussion at the end of section V.

In the next section, the power counting arguments presented above will be used to derive the effective Lagrangian 
which describes the interaction of collinear modes with Glauber exchanges with the medium. While these power counting 
arguments have been derived for an abelian theory, we expect them to remain true in a non-Abelian theory as well.


\section{Effective Lagrangian for Glauber Gluons}


In the preceding section, the momentum scales associated with a new mode which arises in the presence of a medium 
was outlined. These, so called Glauber gluons present a mode that was absent in the derivation of the SCET 
Lagrangian. In what follows, we introduce these modes and construct a new additional effective Lagrangian 
called the Glauber Lagrangian. In this first attempt, the kinetic terms which represent the soft and collinear gluons 
of the SCET Lagrangian, along with their interactions with the collinear quarks will be ignored. The two active 
fields will be the collinear quarks and the Glauber gluons. In principle there will be a similar contribution from Glauber 
interactions with a collinear gluon. While such interactions are not included in the derivations presented in the current paper, these represent a simple extension of the formalism presented in this section.

Before we start the derivation of the effective Lagrangian we comment on the off-shellness of the Glauber gluons. Since for Glauber gluons the product $k^+k^-$ is much less than $\vert \vec{k}_\perp\vert^2$ then the Glauber modes are obviously off-shell degrees of freedom. In principle when one constructs and effective Lagrangian, the degrees of freedom involved have to be on-shell so that one can make use of the classical equations of motions. In our derivation below we do not make any use of the gluon equation of motion and it is only the Dirac equation for a collinear  quark that is utilized. Thus the derivation below should be viewed merely as the limit of the contribution from gluons with arbitrary momentum taken to the Glauber region. This is much similar to the well-known method of regions. In this sense we are deriving actually an effective ``vertex'' between a collinear quark and a Glauber Gluon. This effective vertex could also be obtained on a case by case basis starting from the full QCD amplitudes and then taking the Glauber limit. This is illustrated below for a non-trivial case. The advantage of deriving an effective Lagrangian is mainly the consistent power counting (of the gluon fields and momenta) invoked in the derivation.

In DIS, in the Breit frame, the final state after the hard scattering consists of a hard out-going quark in the $-z$ or simply the ($-$)-direction,  
which interacts with the remnants of the proton (or nucleus) which move in the $+z$ or simply the ($+$)-direction. 
These interactions are dominated by soft grazing scatterings with the fast moving remnants. Some of these interactions 
may be hard as well, resulting in the hard quark going considerably off-shell and radiating a hard gluon. Such processes constitute the 
energy loss of the hard quark and will be ignored in this first attempt. The incorporation of such interactions will necessarily 
involve the reintroduction of both the soft and collinear modes as well as the inclusion of new interaction terms between the 
Glauber modes and these soft and collinear modes. In what follows we focus solely on the soft interactions
 of the hard outgoing quark with the soft glue field generated by the remnants of the struck proton (or nucleus).

Consider a fast-moving quark moving along the ${\bar n}$ direction where ${\bar n}=\frac{1}{\sqrt2}(1,0,0,-1)$.
The $+z$ direction will be denoted by the unit vector, $ n=\frac{1}{\sqrt2}(1,0,0,1)$. 
This quark has large momentum in the $-z$-direction $p^-=n\cdot p\simeq Q$ where 
$Q$ is the hard scale of the process considered. The interaction of this collinear quark with Glauber gluons leads to 
change of the transverse momentum component of the collinear quark while the $p^-$ component remains fixed up to 
${\cal O}(\lambda^2)$. Thus the truly label-changing component is only the transverse one $p_{\perp}$. As such the 
starting point to describe the interaction of this collinear quark field with Glauber gluons
would be, as in SCET~\cite{Bauer:2000ew,Bauer:2000yr}, to extract the label momentum components from the full QCD field. 
\bea
\psi(x)&=&e^{-ip^-x^+}\sum_{\pb}e^{i\pb \cdot \xb} \psi_{{\bar n},\pb}(x).
\label{f1}
\eea

Again as in SCET, we decompose the $\psi_{{\bar n},\pb}$ into a sum of two fields: $\xi_{{\bar n},\pb}+\xi_{ n,\pb}$ where $\xi_{{\bar n},\pb}$ carries the large momentum components in the $-z$ direction while $\xi_{n,\pb}$ carries the small momentum components. The next step is to substitute this decomposition into Eq.~(\ref{f1}) and then substitute the result into the  interaction term in the full QCD Lagrangian. In order to maintain consistent power counting in the effective theory one has to specify the scalings (in terms of $\lambda$) of the relevant quantum fields. For collinear quark the scaling is the same as in SCET, namely $ \xi_{{\bar n},\pb}$ scales as $\lambda$ while $\xi_{n,\pb}$ scales as $\lambda^2$. All derivatives acting on $\xi_{{\bar n},\pb}$ or $\xi_{n,\pb}$ will further suppress the power counting by $\lambda^2$. For the Glauber gluon gauge field the scaling was given in the previous section.

The starting point to obtain the effective Lagrangian is, as in SCET, the full QCD quark sector expressed in terms of 
the fields $\xi_{\bar n}$ and $\xi_n$
\bea
\label{eff0}
{\cal L}_g&=&{\bar \xi}_{{\bar n},\pb}\not\!n(i{\bar n}\cdot D)\xi_{{\bar n},\pb}+{\bar \xi}_{n,\pb} \not\!{\bar n}(n\cdot p{+in\cdot D}) \xi_{n,\pb}\nonumber\\
&&+{\bar \xi}_{{\bar n},\vec{p'}_\perp}(\not\!p_{\perp}+i\not\!D_{\perp})\xi_{n,\pb}+{\bar \xi}_ {n,\vec{p'}_\perp}(\not\!p_{\perp}+i\not\!D_{\perp})\xi_{{\bar n},\pb}.
\eea
We notice that in the last result there are terms that scale as ${\cal O}(\lambda^4)$, ${\cal O}(\lambda^5)$ {and} ${\cal O}(\lambda^6)$.
 We now eliminate the non-dynamical field $\xi_n$ by making use of the tree-level equation of motion
\bea
\label{eqm}
\xi_{n,\pb}&=&\frac{\pbs+{i}\dps}{2{(n\cdot p+in\cdot D)}} \not\!n\xi_{{\bar n},\pb}\,\,.
\eea
We again have kept leading and sub-leading contributions in the Eq.~(\ref{eqm}). 
It is useful to notice the difference between the case of Glauber gluons and collinear gluons. For collinear gluons, the gauge field component in the 
covariant derivative  ($n \x A$) in the denominator of Eq.~(\ref{eqm}) scales the same as $n\cdot p$.  This component of the gauge field, eventually, leads to the presence of the collinear Wilson lines in SCET, as was demonstrated in Ref.~\cite{Bauer:2001ct}. 
However, for Glauber gluons this covariant derivative is suppressed compared to $ n\cdot p$. Therefore, we expand the denominator and get
\begin{equation}
\label{lag}
{\cal L}_g=\sum_{ \pb,{\vec{p'}_\perp}}e^{i({\vec{p}_\perp}-{
\vec{p}'_\perp})\cdot \vec{x}_\perp}{\bar \xi}_{{\bar n},\vec{p'}_\perp}(x)\left[{\bar n}\cdot
iD+(\not\!p_\perp+i\not\!D_\perp)\frac{1}{2n \cdot p}{\left(1-\frac{in\cdot D}{n\cdot p}\right)}(\not\!p_\perp+i\not\!D_\perp)\right]\not\!n\xi_{{\bar n},\vec{p}_\perp}(x)~.
\end{equation}
where higher orders in $\lambda$ have been dropped out. In this article we will only consider Glauber gluons in 
covariant and light-cone ($A^+ = 0$) gauge.
%
In covariant Gauge the leading order Lagrangian is given by
\bea
{\cal L}_g={\bar \xi}_{{\bar n},\pb}\left[i{\bar n}\cdot D+\frac{p_\perp^2}{2n\cdot p}\right]\not\!n \xi_{{\bar n},\pb}
\eea
where both terms in the square brackets are of order $\lambda^2$.
In light-cone gauge the leading order interaction Lagrangian is given by
\bea
{\cal L}_g={\bar \xi}_{{\bar n},\vec{p'}_\perp}\left[\frac{g_s(\not\!p'_{\perp} {\not\!A}_\perp+{\not\!A}_\perp \not\!p_\perp)+g_s^2{\not\!A}^2_{\perp}}{2n\cdot p}\right]\not\!n\xi_{{\bar n},\pb}.
\eea

\begin{figure}[htbp!]
 \begin{center}
\includegraphics{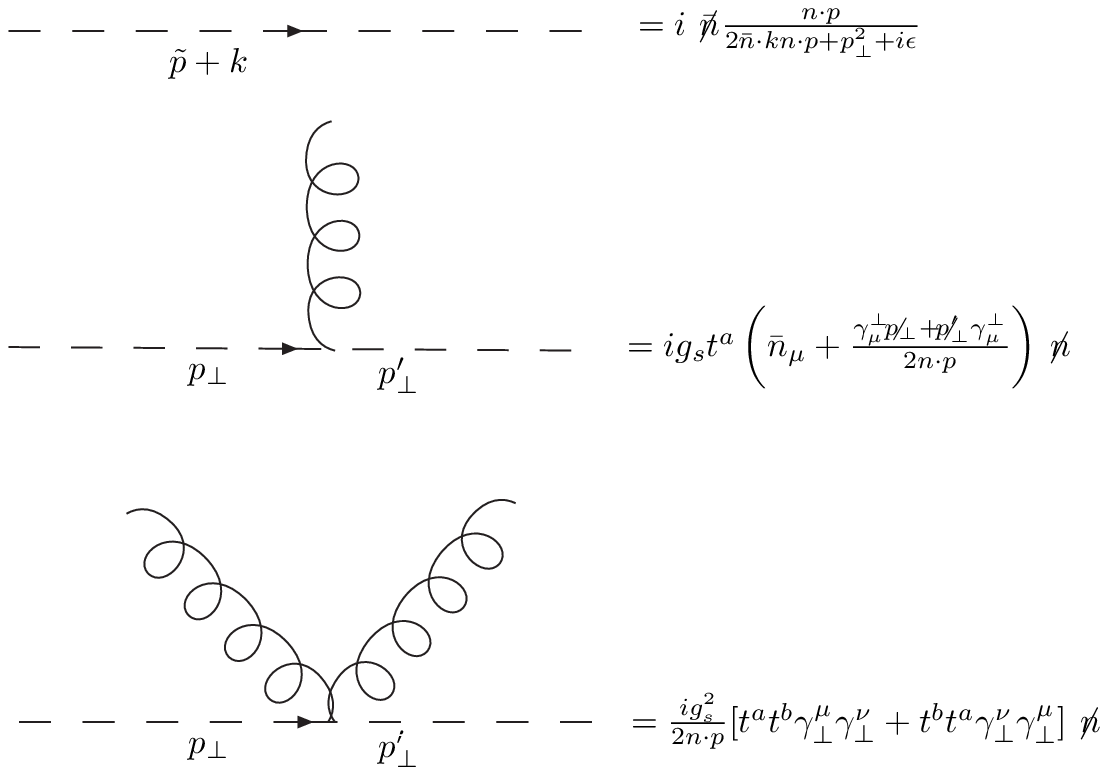}
 \end{center}
 \caption{Feynman rules for a collinear quark interacting with Glauber gluons. In the second diagram with a single 
gluon vertex, the first term is the contribution from the covariant gauge Lagrangian while the second term is the 
leading contribution from the light-cone gauge Lagrangian. The third diagram is only required for calculations in 
light-cone gauge.}
\label{GLF1}
\end{figure}

The Feynman rules derived from the effective Lagrangians above are given in Fig.~\ref{GLF1}. 
As a first simple test of these rules we compute the 
amplitude for the case of two Glauber gluons attached to a collinear quark line and compare with the amplitude 
obtained from full QCD. We show that the effective theory exactly reproduces the 
full QCD result at the level of the amplitudes of Feynman diagrams. 

\begin{figure}[htbp!]
\resizebox{4in}{2in}{\includegraphics[3in,8in][7in,10in]{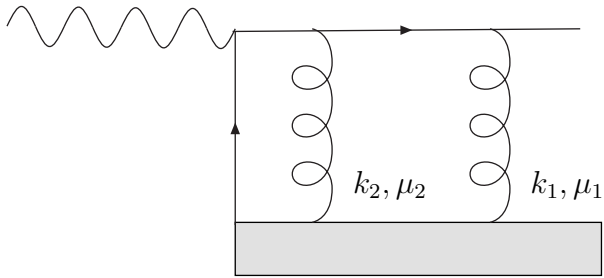}} 
    \caption{Final state interactions in DIS: two gluon exchange.}
    \label{FS}
\end{figure}

Let us consider the Feynman diagram given in Fig.~\ref{FS} where two Glauber gluons are attached to the collinear quark field.
In full QCD, the amplitude reads
\bea
\label{app1}
\label{full}
I = - (ig_s)^2 \int { \frac{d^4 k_1 }{ (2 \pi)^4}\frac{d^4k_2}{(2\pi)^4}} {\bar u}(p) \frac{\not\!A(k_1)[\not\!p-\not\!k_1]\not\!A(k_2)[\not\!p-\not\!k_1-\not\!k_2] }{[(p-k_1)^2+i\varepsilon][(p-k_1-k_2)^2+i\varepsilon]},
\eea
where ${\bar u}(p)$ represents the Dirac spinor for an outgoing quark and $p$ has no transverse momentum. 
The scaling of the quark spinor will be ignored in the following, as it plays no role in the remaining discussion.   
When expanding the 
numerator in Eq.~(\ref {app1}) one should invoke the same power counting for the gluon gauge fields 
and the momenta as the one used in deriving the effective theory. In light-cone gauge $A^+=0$ and by making 
use of the Dirac equation [${\bar u}(p)\not\!p=p^-{\bar u}(p)\gamma^+=0$], the leading contribution is,
\bea
\label{app2}
{\cal J} &=&  -(ig_s)^2 \int { \frac{d^4 k_1 }{ (2 \pi)^4} \frac{d^4k_2}{(2\pi)^4}}{\bar u(p)}\frac{ \not\!A_\perp(k_1)[2k_1^+p^-\not\!A_\perp(k_2)
\mbx\, + \not\!k_{1\perp}\not\!A_\perp(k_2)(\not\!k_{1\perp}+\not\!k_{2\perp})]}{[(p-k_1)^2+i\varepsilon][(p-k_1-k_2)^2+i\varepsilon]} .
\eea
We notice that each contribution in the square bracket scales as $\lambda^4$ and subleading terms have been dropped out.

In the effective theory (and again working in light-cone gauge) there are two Feynman diagrams that contribute. One (denoted below as ${\cal J}^{(1)}$) comes from two Glauber gluons attached at different points. The other contribution comes from the vertex of two glauber gluons attached at the same point. It is denoted by 
${\cal J}^{(2)}$.
Using the Feynman rules given in Fig.~\ref{GLF1} we obtain the first contribution from the effective theory as, %
\bea
{\cal J}^{(1)}&=& 
- (ig_s)^2 \int \frac{d^4 k_1 }{ (2 \pi)^4}\frac{d^4k_2}{(2\pi)^4}  ~{\bar \xi}_{\bar n}
\frac{\not\!A_{\perp}(k_1)\not\!k_{1\perp}[\not\!A_{\perp}(k_2)(\not\!k_{1\perp}+\not\!k_{2\perp})+\not\!k_{1\perp}\not\!A_{\perp}(k_2)] }{[2p^-k_1^+ +\vert \vec{k}_{1\perp}\vert^2-i\varepsilon][2p^-(k_1^+ +k_2^+)+ \vert  \vec{k}_{1\perp}+\vec{k}_{2\perp}\vert^2-i\varepsilon]}.
\eea
The second contribution from the effective theory is given as, 
\bea
{\cal J}^{(2)}&=& - (ig_s)^2\int \frac{d^4 k_1 }{ (2 \pi)^4}\frac{d^4k_2}{(2\pi)^4} ~{\bar \xi}_{\bar n}\frac{\not\!A_\perp(k_1)\not\!A_\perp(k_2)}{[2p^-(k_1^++k_2^+)+\vert \vec{k}_{1\perp}+\vec{k}_{2\perp}\vert^2-i\varepsilon]}.
\eea
In the above result we have considered only one contribution where the color and Lorentz indices are held fixed 
in the Feynman rule for the two-gluon vertex. The other contribution gives an identical result as ${\cal J}^{(2)}$ 
upon integrations over $k_1$ and $k_2$.
It is can be easily verified that ${\cal J}={\cal J}^{(1)}+{\cal J}^{(2)}$ thus confirming that the effective theory 
reproduces the full QCD result at the level of the amplitudes of the relevant Feynman diagrams.
The case of one-gluon attachment is trivial and one can easily verify that the effective Lagrangian 
also gives the same result as the one in full QCD.

In what follows, we investigate two physical applications of the derived effective Lagrangian: The transverse 
broadening experienced by hard jets in DIS on a large nucleus and the transverse momentum dependent 
parton distribution function (TMDPDF) in a nucleon. In the first case, a final physical cross section will be computed 
and arguments on the enhancement of power corrections by large lengths in nuclei will be forwarded; hence, this 
application contains arguments beyond those used to derive the effective Lagrangian. However, the scaling 
of the momenta of the gluons will always lie within the strict boundary prescribed by the Glauber Lagrangian. 
It may not come as a surprise that the Glauber gluons which lead to the transverse broadening of hard jets, also 
play a principal role in the construction of the gauge invariant TMDPDF.


\section{Application I: Transverse broadening in large nuclei}


A straightforward application of the effective Lagrangian derived in Sec. III, is to the process of 
jet broadening in dense matter. As a specific example, we consider the process of jet broadening in 
Deep-Inelastic scattering in large nuclei. A virtual photon with momentum $q=[Q^2/2n \x q,n \x q, 0,0]$ 
is incident on a large nucleus ($A$) with momentum $A p\x \bar{n}$ where $p\x \bar{n}$ is the mean momentum 
of a nucleon. In the remaining section, we will refer to $q\x  n$ as simply $q^-$ and $p\x \bar{n}$ as simply $p^+$.

We compute the cross section for the semi-inclusive production of a  hard jet in the 
final state with a net transverse momentum $\vl_\perp$ with respect to the direction of the 
virtual photon, i.e., 
\bea
e(L_1) + A(p) \longrightarrow e(L_2) + J(\vl_\perp) + X .
\label{chemical_eqn}
\eea
In the frame chosen, the Bjorken variable is defined as $x_B = Q^2/(2p^+ q^-)$. The differential 
cross section may be decomposed into a leptonic and a hadronic part as, 
\bea
\frac{E_{L_2} d \sigma } {d^3 L_2 d^2 l_\perp } &=&
\frac{\A_{EM}^2}{2\pi s  Q^4}  L_{\mu \nu}  
\frac{d W^{\mu \nu}}{d^2 l_\perp}. \label{LO_cross}
\eea
\nt
where $s = (p+L_1)^2$ is the total invariant mass of the lepton nucleon 
system. The leptonic tensor may be expressed as,
\bea 
L_{\mu \nu} = \frac{1}{2} \tr [ \f L_1 \g_{\mu} \f L_2 \g_{\nu}].
\eea
The initial state of the incoming nucleus is defined as $| A; p \rc$.
The  general final hadronic or partonic state is defined as 
$| X \rc $.
As a result, the semi-inclusive hadronic tensor may be defined as 
\bea 
W^{\mu \nu}\!\!\!\!&=& \!\!\!\! \sum_X \!\!(2\pi^4) 
\kd^4 (q\!+\!P_A\!-\!p_X ) 
\lc A; p |  J^{\mu}(0) | X  \rc \lc X  | J^{\nu}(0) | A;p \rc  
= 2 \mbox{Im} \left[  \int d^4 y e^{i q \cdot y } \lc A;p | J^{\mu} (y) J^{\nu}(0) | A;p \rc \right],  \label{W_mu_nu}
\eea
where the sum ($\sum_X$)  runs over all possible hadronic states and $J^{\mu}$ is the 
hadronic electromagnetic current i.e., $J^{\mu} =  Q_q \bar{\xi}_{\bar{n}} \g^\mu \xi_n$, where $Q_q$ is the 
charge of a quark of flavor $q$ in units of the positron charge $e$. 
It is understood that the factors of the electromagnetic coupling constant have already been extracted and 
included in Eq.~\eqref{LO_cross}.
The leptonic tensor will not be discussed further. The focus in the remaining 
shall lie exclusively on the hadronic tensor.

In a full QCD calculation of Eq.~\eqref{W_mu_nu}, one computes the hadronic 
tensor, order by order, in the strong coupling. This leads to the introduction of a 
variety of processes leading to a modification of the structure of the jet. 
Such processes include radiative branchings, flavor changes of propagating partons, as 
well as transverse diffusion of the partons in the shower which ensues from the 
quark produced in the hard scattering. In this article, we will focus solely on the 
processes which lead to the transverse momentum diffusion or transverse broadening 
of the produced hard quark. 

\begin{figure}[htbp]
\resizebox{4in}{3in}{\includegraphics[0in,0in][8in,6in]{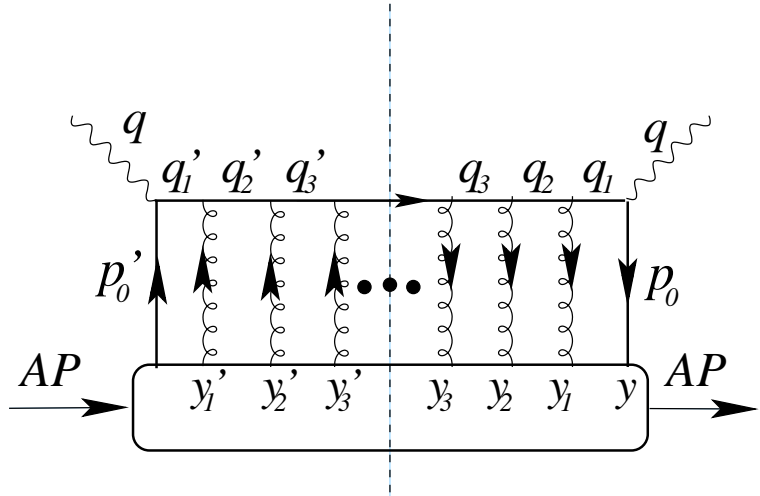}} 
    \caption{An order $n$ diagram which contributes solely to transverse broadening. }
    \label{trans_broad}
\end{figure}

In Ref.~\cite{Majumder:2007hx}, the leading contributions to transverse broadening 
without induced radiation, at all orders in coupling,  were identified as those 
of Fig.~\ref{trans_broad}. These diagrams depict processes where the propagating 
parton engenders multiple scattering off the glue field inside the various nucleons 
through which it propagates. However, scatterings do not change the small off-shellness 
of the propagating parton; as a result, large transverse momentum radiations do not occur. Using simple kinematics, the 
relation between the momentum components of the glue field $k_i$ may be surmised by 
insisting that the off-shellness of the $i+1^{\rm th}$ quark line be of the same order as the $i^{\rm th}$ 
line, 
\bea
(p + k_i)^2 = p^2 + k_i^2 + 2 p^+ k_i^-  + 2 p^- k_i^+ - 2 {\vp}_\perp \x { \vec{k}^i }_\perp .
\eea
Insisting that $(p + k_i)^2 \sim p^2 \sim \lambda^2Q^2$ and given the known scaling of 
the quark momenta (i.e., $p^+ \sim \lambda^2 Q, p^- \sim Q, \vp_\perp \sim \lambda Q$), we obtain that 
$\vk^i_\perp \sim \lambda Q$, $k_i^+ \sim \lambda^2 Q$ and $k_i^-$ may scale with a range of 
different choices $Q, \lambda Q, \lambda^2 Q $ etc. The first two cases for the scaling of $k^-$ 
represent gluons which are emanated with large $(-)$-momentum from a nucleon moving 
with large $(+)$-momentum. The number of such gluons must be vanishingly small. The 
first non-trivial population of gluons emanating from a nucleon moving with a large $(+)$-momentum, 
are those which scale as $k \sim [\lambda^2, \lambda^2, \lambda]$, which essentially 
constitute the Glauber sector.

Using the Feynman rules derived for Glauber gluons in section 2, the leading component 
of $n^{\rm th}$ order diagrams such as those of Fig.~\ref{trans_broad} may be 
expressed as, 
\bea
W^{\mu \nu} &=& \int d^4 y \frac{d^4l }{(2\pi)^4}  
\prod_{i=0}^{n-1} \prod_{j=0}^{n^\p-1} \left\{ d^4 y_{i+1} d^4 y_{j+1}^\p 
\frac{d^4 k_{i} d^4 k_j^\p }{(2 \pi)^8} \right\} g^{n+n^\p}   \nn \\
\ata   \lc A;p | \xi_{{n}}(0) \bar{\xi}_{{n}}(y) \g^\mu  
\prod_{i=1}^n \left[ \frac{ 2 q^- }{2q^- k_i^+ - |{\vk^{\,i}}_\perp|^2  - i \e}   \right] 
 \! 2l^-   2\pi \kd(2 l^+ l^- - l_\perp^2)  
\prod_{j=n^\p}^1 \left[ \frac{2q^-}{2q^- {k_j^\p}^+ -  |\vec{k^{\p}}^j_\perp|^2 + i \e} \right]
t^{a_i} A^+_{a_i}(y_i) \nn \\
\ata t^{a^\p_j} A^+_{a^\p_j}(y^\p_j) \g^{\nu} | A;p \rc 
e^{ -i \sum_{i=1}^{n-1} k_i \x y_i   }
e^{i \sum_{j=1}^{n^\p-1}   k^\p_j \x y^\p_j  } 
 e^{\left[ -i y_n \x \left\{ l - \left(q + \sum_{i=0}^{n-1} k_i 
\right) \right\} \right]} 
e^{\left[i y^\p_{n^\p} \x \left\{ l - \left(q + \sum_{j=0}^{n^\p - 1} 
k^\p_j \right) \right\} \right]},  \label{W_mu_nu_general} 
\eea
where, it is understood that $y^\p_0$ is the origin and $y^0 \equiv y$. 
In the equation above, the gauge fields have been expressed in coordinate space.
At this point an $n^{\rm{th}}$ momentum may be introduced, via 
\begin{equation}
1 = \int d^4 k_n  \kd^4 \left( l - \sum_{k=0}^{n} k_k - q \right). \label{pn_delta}
\end{equation}
\nt
This leads to a considerable simplification of the phase factors. 
The complete absence of  the ($-$) components of the momentum, from all expressions except for the 
phase factors allows for the $k^-$ and ${k^\p}^-$ integrations to be done, resulting in the localization of the 
process on the negative light-cone.

The integrals over the momenta $k_i^+, {k^\p}_j^+$ may be re-expressed in terms of momentum fractions, \tie,
\bea
Q^2 = 2x_B p^+ q^-\,; \,k_i^+ = x_i p^+  \,;\, {k^\p}^+_j = x^\p_j p^+ \label{long_def}  \\
\slm_{k=0}^{i} 2\vk_\perp^i \cdot \vk_\perp^k + |\vk_\perp^i|^2 = 2x^i_D p^+q^-;  \label{i_perp_def} \\
\slm_{l=0}^{j} 2 {\vec{k^\p}}_\perp^j \cdot {\vec{k^\p}}_\perp^l + |{\vec{k^\p}}_\perp^j|^2 = 2 {x^\p}^j_D p^+q^- . \label{j_perp_def}
\eea
Integrating over all the $x_i$ and $x^\p_j$ momentum fractions by contour integration, we obtain the 
much simplified form of the hadronic tensor, 
\bea 
W^{\mu \nu } &=& g^{n + n^\p} \int \frac{d^2l_\perp }{(2\pi)^2}  \prod_{i=0}^n d y_i^- d^2 y_\perp^i 
\prod_{j=1}^{n^\p} d {y^\p}_j^- d^2 {y^\p}^j_\perp 
\int \prod_{i=0}^n \frac{ d^2 k^i_\perp}{(2\pi)^2} 
\prod_{j=0}^{n^\p - 1} \frac{d^2 {k^\p}^j_\perp} { (2\pi)^2} 
(2\pi)^2 \kd^2 ( \vec{l}_\perp - \vec{K}_\perp )  \nn \\
\ata \hf \left( g^{\mu - } g^{\nu + } + g^{\mu +} g^{\nu -} - g^{\mu \nu}  \right) 
 e^{-ix_B p^+ y^-} \prod_{i=0}^n e^{-ix_D^i p^+ y_i^- } 
e^{i \vk^{\,i}_\perp \x  \vec{y}^{\,i}_\perp  } 
 \prod_{j=0}^{n^\p} e^{i {x^\p}_D^j p^+ {y^\p}_j^- } 
e^{-i\vec{k^\p}^j_\perp \x  \vec{y^\p}^j_\perp  } \nn \\
\ata \prod_{i=n}^1 \h ( y_i^-   - y_{i-1}^- ) 
\prod_{j = n^\p}^1 \h ( {y^\p}_j^-  -  {y^\p}_{j-1}^- ) \nn \\
\ata \lc A; p | \bar{\xi}_{{n}} (y^-,y_\perp) \g^+ \xi_{{n}}(0) 
\tr \left[ \prod_{i=1}^{n} t^{a_i}  A_{a_i}^+ (y_i^-,\vec{y}^{\,i}_\perp) 
 \prod_{j=n^\p}^{1} t^{a_j} A_{a_j}^+ ( {y^\p}_j^-, \vec{y^\p}^j_\perp ) \right]  | A;p \rc.
\label{W_mu_nu_simple}
\eea
The expression derived above has so far been a direct application of the Feynman rules derived 
in the preceding section. 
Hitherto, no assumption 
regarding the nature of the nuclear state has been made. As a result the nuclear or nucleon scale of 
$\mu Q$ has also not appeared in any of the expressions. 
However, the hadronic tensor in Eq.~\eqref{W_mu_nu_simple} 
and any resulting transverse broadening will, ultimately,  depend on the expectation of the ($n+n^\p+2$)-parton operator as 
indicated in the last line of Eq.~\eqref{W_mu_nu_simple}.
To proceed further, approximations regarding the expectation of this partonic operator will have to be made. 
In these approximations, the in-medium scale $\mu Q$ will appear. However, as we will show, the final transverse 
broadening will turn out to be independent of this scale under certain assumptions.

Following standard treatments, we approximate the  nucleus as a weakly interacting homogeneous gas of nucleons. 
Such an approximation is only sensible at very high energy, where, due to 
time dilation, the nucleons appear to travel in straight lines almost independent of each other 
over the interval of the interaction of the hard probe. All forms of correlators  
between nucleons (spin, momentum, etc.) are assumed to be rather suppressed. 
As a result, the expectation value of 
the $n+n^\p+2$ operators in the nuclear state may be decomposed as 
\bea
\left \langle A;p \left| \bar{\xi}_{\bar{n}}(y^-,\vec{y}_\perp)\g^+ \xi_{\bar{n}}(0) \prod_{i=1}^{n+n^\p} A^{+}_{a_i}(y_i) \right| A; p\right \rangle 
\!\!\!&=&\!\!\! C^A_{p_0,p_2,...p_n}  \lc p_0 |  \bar{\xi}_{\bar{n}}(y^-,\vec{y}_\perp)\g^+ \xi_{\bar{n}} | p_0 \rc 
\! \prod_{i=1}^{(n+n^\p)/2} \!\!\lc p_i | A^{+}_{a_i}(y_i)  A^{+}_{a^\p_i} (y^\p_i)  | p_i \rc  , \label{factorized_matrix_element}
\eea
\nt 
where, the factor $C^A_{p_0,p_1,\ldots, p_n}$ represents the correlations between the $(n+n^\p)/2$ 
nucleons which interact with the propagating parton. In the decomposition above, we have restricted 
at most two parton operators per nucleon, insisting that any larger number of operators is suppressed.
Note that this is only true outside the saturation regime~\cite{Mueller:1989st,McLerran:1993ni}. The 
decomposition performed above, also restricts $n=n^\p$. 
The choice of gluon operators per nucleon is also special to the case of transverse broadening: 
The maximum broadening is obtained when one gluon operator from the amplitude 
is paired with one from the complex conjugate. This reason for this is immediately understood with further 
simplifications on each gluon pair (written with spin and color indices suppressed), 
\bea
\mbx && \int  d^2 y^i_\perp  d^2 {y^\p}^j_\perp \lc p | A^{+}  (\vec{y}^i_\perp )  A^{+}  ( \vec{y^\p}^j_\perp ) | p \rc 
 \exp \left[ {-i x_D^i p^+ y_i^-}  + { i \vk^{\,i}_\perp \x \vec{y}^{\,i}_\perp}   
+ {i {x^\p}_D^j p^+ {y^\p}_j^-}  { - i \vec{k^\p}^j_\perp \x \vec{y^\p}^j_\perp} \right] \nn \\
&=& (2\pi)^2 \kd^2( {\vk}^{\,i}_\perp - {\vec{k^\p}} ^j_\perp )  \int d^2 y_\perp  e^{-i x_D^i p^+ ( y_i^-  -  {y^\p}_j^- )} 
e^{ i \vk_\perp \x \vec{y}_\perp} \lc p | A^{+} (\vec{y}_\perp/2 )  A^{+}  ( - \vec{y}_\perp/2 ) | p \rc , \label{two_gluon_cor}
\eea 
where, $\vec{y}_\perp$ is the transverse gap between the two gluon insertions and 
$\vk_\perp = (\vk^{\,i}_\perp + \vec{k^\p}^j_\perp)/2$. 
The physics of the 
above equation is essentially the transverse translation symmetry of the two gluon correlator 
in a very large nucleus. 
This is then used to equate the transverse 
momenta emanating from the two gluon insertions. Thus, if the two operators were both chosen from the amplitude or 
the complex conjugate, then the momenta brought in by one gluon operator would be immediately taken out by the 
other and, as a result, the combination of the two operators will lead to no net transverse broadening. 
The integration above, also simplifies the longitudinal phase factors which now depend solely on the difference of the 
longitudinal positions of the two gluon insertions.

Further simplifications are introduced by Taylor expanding the transverse momentum dependent delta function, as 
\bea
 \prod_{i=1}^n \frac{\prt^2  }{2! \prt^2  k_\perp^i} 
\left.  \kd^2(\vl_\perp + \sum \vk^{\,i}_\perp) \right|_{\vk^{\,i}_\perp = 0} \prod_{i=1}^n|\vk^{\,i}_\perp|^2,
\eea
and combining the $|\vk^{\,i}_\perp|^2$ with the expectation of the two gluon operator
 $\lc p_i | A^{+}_{a_i}(\vec{y}^i_\perp/2)  A^{+}_{a^\p_i} (-\vec{y}^i_\perp/2)  | p_i \rc $ to convert these into 
the expectation of field strengths in the nucleon 
$\lc p_i | F^{+ \perp }_{a_i}(\vec{y}^i_\perp/2)  F^{+ \perp}_{a^\p_i} (-\vec{y}^i_\perp/2)  | p_i \rc$. 
The meaning of this decomposition of the transverse momentum delta function is the retention of 
solely the leading twist part of each of the two-point correlators in each nucleon. Higher powers of a given
transverse momentum will necessarily lead to higher transverse moments of the two gluon operator.
One further assumption regarding the two point function of Eq.~\eqref{two_gluon_cor}, 
due to color confinement, leads to a constraint on the two longitudinal $y^-$ integrations (ignoring color and spin indices), 
\bea
\int d  y^- d {y^\p}^- \lc p | F  (y^- )  F ({y^\p}^- ) | p \rc  \simeq \int d y^-  \lc F F \rc y^-_{conf}, 
\eea
where, $\lc FF \rc$ is the gluon expectation in a nucleon and $y^-_{conf}$ is the confining distance.

Each such integral yields a factor of $L^- \sim A^{1/3} \sim 1/\lambda$ from the unconstrained $y^-$ integration.  
The equating of the pairs of transverse momenta that appear in each two-gluon correlation, as 
well as the relation between the longitudinal momenta from the $\h$-functions in Eq.~\eqref{W_mu_nu_simple}, 
require that the largest transverse momentum broadening and largest length enhancement arises from the terms 
where the gluon correlations are built up in a mirror symmetric fashion, \tie, where the gluon insertion at $y^i$ is 
contracted with that at ${y^\p}^i$.

Averaging over the spins and colors of the two point functions in each nucleon, 
the remaining $n$ longitudinal position integrals for the gluon insertions may be simplified as 
\bea
\int \prod_{i=1}^{n} dy_i^-  \h(y_i^- - y_{i-1}^-) = \frac{1}{n!} \int \prod_{i=1}^n dy_i^- .
\eea
\nt
Invoking the above simplifications, the leading length enhanced contribution at order $2n$ to the differential hadronic 
tensor is obtained as,
\bea 
\frac{d^2 W_n^{\mu \nu}}{d^2 l_\perp} &=& C^A_{p_0,\ldots, p_n}W_0^{\mu \nu} \frac{1}{n!} \left[ \{\nabla^2_{l_\perp}\}^n \kd^2 (\vec{l}_\perp) \right] 
\left[ \frac{\pi^2 \A_s}{2N_c} L^- \int \frac{dy^-}{2\pi} \lc p | {F^a}^{+ \A} {F^a}_{\A,}^{\,\,\, +}| p \rc \right]^n.  \label{W_n}
\eea
\nt
where $W_0^{\mu \nu}$ is the leading order transverse momentum integrated hadronic tensor, given as 
\bea
W_0^{\mu \nu} &=& 2 \pi  [ g^{\mu -} g^{\nu +} + g^{\mu +} g^{\nu -} - g^{\mu \nu} ]  \sum_q Q_q^2 f_q(x_B) ,  \label{W_0}
\eea
where, the expectation of the two quark operator in Eq.~\eqref{factorized_matrix_element}, leads to the quark structure 
function in the equation above.

There remains the overall coefficient $C^A_{p_0,\ldots, p_n}$ which contains the 
weak correlations between the various struck nucleons. A study of such correlations in Refs.~\cite{Osborne:2002st,Majumder:2008jy}
revealed that a simple factorized form such as  $C^A_{p_0,\ldots, p_n} = C^A_{p_0}  (\rho/2p^+)^n$, where $\rho$ is the 
density of nucleons in a nucleus, is not completely inappropriate. Using this simple form one may sum over all $n$, i.e., over 
multiple scatterings of the quark in the nucleus, to obtain the resummed equation, 
\bea
\!\!\!\!\!\!\!\!
\frac{d^2 W^{\mu \nu }}{d^2 l_\perp} = e^{ (D L^-) \nabla^2_{l_\perp} } 
\frac{d^2 W_0^{\mu \nu }}{d^2 l_\perp},  \label{resummed}
\eea
\nt
where, $d^2W_0^{\mu \nu}/d^2 l_\perp = W_0^{\mu \nu} \kd^2 (\vl_\perp) $  , and the constant 
$D$ is given as, 
\bea
D &=& \frac{\pi^2 \A_s}{2 N_c} \rho \int 
\frac{d^3 y  d^2 k_\perp}{(2\pi)^3 2 p^+} \lc p | {F^a}^{+ \A}(y) {F^a}_{\A,}^{\,\,\, +}(0)| p \rc 
%
\exp \left\{-i \left( \frac{|\vk_\perp|^2}{2q^-}  y^- -  \vk_\perp \x \vec{y}_\perp \right) \right\} .
\label{D}
\eea
It is this constant $D$ which controls the broadening experienced by the hard jet in the extended nucleus.

As shown in Ref.~\cite{Majumder:2007hx}, Eq.~\eqref{resummed} is a solution of the two dimensional transverse momentum 
diffusion equation, where the initial condition may be taken as a $\kd$-function in transverse momentum. 
Taking moments of the 
solution of the diffusion equation, we obtain the total transverse momentum squared acquired by the hard quark after 
traversing a length $L^-$ in the nucleus as given by the simple relation,
\bea
\lc k_\perp^2 \rc_{L^-} &=& 4 D L^-. \label{p_T_and_L}
\eea
The reader will note that we have used a two dimensional delta function as the input to the diffusion equation. This is an approximation 
to a very peaked distribution and 
one may use any other input distribution as well. The net extra broadening experienced by the input distribution is 
given by Eq.~\eqref{p_T_and_L}. Given that the initial parton is an SCET mode, the transverse momentum is of the 
order of $|\vk_\perp|^2 \sim (\lambda Q)^2$. As a last step, we will demonstrate that the broadening obtained from 
multiple scattering in the large nucleus is of this order in power counting and thus one may continue to think of an 
SCET mode propagating in the extended medium.

The power counting of net transverse momentum squared may be easily estimated from counting powers of 
$\lambda$ and $\mu$ in the expression for $D$ in Eq.~\eqref{D}. As the expression is frame independent it will be evaluated in the Breit frame. In this frame, the $z$-component of the length of the nucleon ($\sim 1/(\mu Q)$) is contracted to a length of order $1/Q$; hence the nucleon density $\rho$ scales as
\bea
\rho &=& \frac{1}{V} \sim \mu^2 Q^3.
\eea 
The dimension of the nucleon ket is obtained from the standard normalization of the on-shell nucleon state, given as,
\bea
\lc p | q \rc = (2 \pi)^3 2 p^+ \kd (p^+ - q^+) 
\kd^2 (\vp_\perp - \vq_\perp)   \Longrightarrow   | p \rc \sim ( \mu Q )^{-2}.
\eea
The $F^{+ \perp}F^{+ \perp}$ correlator scales as $ ( \lambda^3 Q^2 )^2 $ from the standard Glauber scaling rules for the 
transverse momentum and the vector-potential. The enhanced length in the nucleus may be expressed in terms of the nuclear parameter 
$A^{1/3}$ and the length of a nucleon $l_N$, as 
\bea
L^- = A^{\frac{1}{3}} l_N \sim \frac{A^{\frac{1}{3}}  }{ Q}.
\eea

Substituting the above relations in Eq.~\eqref{D}, and noting that the $ y^-$ and the $\vec{ y}_\perp$ coordinates 
are conjugate to the $|k_\perp|^2/2q^-$ and $\vec{p}_\perp$ momenta,   
yields the $\lambda$ power counting of the net transverse momentum 
squared picked up by the hard parton as, 
\bea
\lc p_\perp^2 \rc_{L^-}  \sim A^{\frac{1}{3}}  \lambda^4 Q^2, \label{A_scaling}
\eea
independent of the medium scaling parameter $\mu$. The broadening is rather small in an object the size of a nucleon, 
but may get enhanced in large or dense media.
As a result, for small nuclei where $A^{1/3} \ll \lambda^{-2}$ one may ignore this extra effect of final state multiple 
scattering. For nuclei, where $A^{1/3} \sim \lambda^{-2}$, or the medium has a very large gluon density we 
may obtain a broadening which is comparable to the jets inherent transverse momentum. In this case, 
we are in the ``SCET-Glauber'' region, where the derived effective 
Lagrangian in this paper may be used in combination with the SCET Lagrangian to understand the interaction 
of hard jets in dense media. For nuclei where $A^{1/3} \gg \lambda^{-2}$, the Glauber modes will broaden the 
propagating jets beyond the scaling assumed in the derivation of the SCET Lagrangian and a different set of 
effective theories will need to be constructed.

There is an unknown quantity that has been invoked a number of times in the discussion above: the inherent gluon 
density. This is the density of ``small $x$'' gluons that emanate from the current density in the medium and interact 
with the hard jet. The number of such gluons is a dimensionless quantity and thus difficult to estimate in a power counting 
calculation. The number of such gluons may in general also depend on the media in question. It is well known 
that the number of such gluons may become rather large at high energies and may thus lead to considerable 
broadening of hard jets.
While the application in this section has focused on the broadening of jets in nuclei, the factorization 
properties afforded by SCET bode well for the applicability of this theory to jet broadening in dense Quark-Gluon-Plasmas (QGPs) 
created in heavy-ion collisions. While QGPs have been estimated to be from 10 to 100 times denser than nuclear matter, 
their lifetimes are rather short and the majority of jets propagate rather short distances in the densest part of such environments. 
Given these experimental considerations, we expect the derived effective theory in combination with SCET to have 
wide applicability.


\section{Application II: TMDPDF}


Inclusive hard scattering processes like DIS can be factorized, into perturbatively calculable short-distance quantities convoluted with non-perturbative long-distance quantities \cite{ster}. The latter quantities are the familiar Feynman PDFs. For semi-inclusive processes where a single hadron is observed in the final state with a given transverse momentum then it is the TMDPDF that enters into the factorization formula for the cross-section. More details can be found in \cite{col,fac}. The TMDPDF have been introduced long time ago in \cite{Collins:1981uw}, as,
\bea
f(x,\vec{k}_\perp)&=&\frac{1}{2}\int \frac{d\xi^-d^2\vec{\xi}_\perp}{(2\pi)^3}e^{-i(\xi^-k^+-\vec{\xi}_\perp\cdot \vec{k}_\perp)}
\times \langle P\vert {\bar \psi}(\xi^-,\vec{\xi}_\perp)L^\dagger_{\vec{\xi}_\perp}(\infty,\xi^-)
\times \gamma^+ L_0(\infty,0)\psi(0,\vec{0}_\perp)\vert P \rangle ,
\eea
where $L$ is path-ordered gauge link along the light-cone in the ($-$) direction, i.e.,
\bea
L_{\vec{\xi}_{\perp}}(\infty,\xi^-)=P \exp \left(ig_s\int_{\xi^-}^\infty  d\xi^- A^+(\xi^-,\vec{\xi}_\perp)\right).
\eea
The Wilson line $L_{\vec{\xi}_\perp}$ has a well-known origin: It comes from radiation of gluons which are collinear  to the incoming parton. In SCET these Wilson lines are also the familiar collinear Wilson lines $W$ \cite{Bauer:2001ct}. In this work we will not discuss further the emergence of those Wilson lines as they are not related to Glauber gluons.

The TMDPDF is a physical quantity  and thus has to be gauge-invariant under arbitrary gauge transformation. However the above  definition is gauge-invariant only in the set of non-singular gauges like covariant gauges where the gluon field vanishes at $\xi^-=\infty$. In singular gauges like light-cone gauge (with $A^+=0$) the gluon field (specifically the transverse components $A_\perp$) does not vanish at $\xi^-=\infty$ and a gauge transformation performed with $A_\perp(\xi^-=\infty)$ will generate a non-vanishing phase that is not compensated by any gauge link. Thus the above definition of the TMDPDF has to be modified by introducing an additional  gauge link formed from the transverse components $A_\perp$:
\bea
L_{\xi^-=\infty}(\vec{\xi}_\perp,\vec{0}_\perp)=P\exp\left(ig_s\int_0^\infty d\vec{\xi}_\perp \cdot \vec{A}_\perp(\xi^-=\infty,\vec{\xi}_\perp)\right),
\eea
 where the line integral in the transverse plane can be performed in arbitrary direction. 
The above observations were first made in Ref~.\cite{ji1}. 

The important question that arises is what kind of interactions, say in DIS, build up this gauge link. This question was answered in the work of Belitsky, Ji and Yuan (BJY) \cite{bjy}. There it was shown that the final-state interactions between the struck quark and the remnants of the incoming proton are responsible for the appearance of this gauge link. Those final-state interactions are mediated by Coulomb gluons that carry mainly a momentum in the transverse direction. In the next section we will check whether these gluons are Glauber gluons or not. It is important to verify this issue as the final state interactions are responsible for many physical effects like single-spin asymmetries and shadowing \cite{bro1,bro2,bro3}. Any attempt to formulate an effective field theoretic approach to study such effects has to start from identifying the relevant momentum modes that mediate the interactions.

\begin{figure}[htbp!]
\begin{center}
\resizebox{4in}{2in}{\includegraphics[3in,8in][7in,10in]{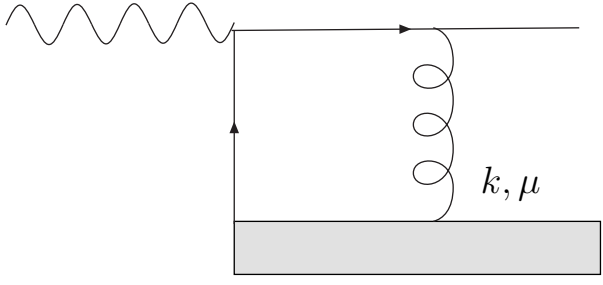}} 
    \caption{Final state interactions in DIS: one gluon exchange.}
    \label{FSI}
 \end{center}
\end{figure}

We start by briefly reviewing the work of BJY by considering the Feynman diagram given in Fig.~\ref{FSI}. A quark propagator with momentum $p-k$ has a denominator:
$-2p^-k^++k^2$ where the quark has essentially large momentum in the $-z$-direction. The integral over the gluon momentum $k$ gets contributions from
a vanishing denominator. This could happen if (1) $k$ is collinear to the outgoing quark. In this case both terms in the denominator scale as $\lambda^2$ or (2) $k$ is soft which means $k^2\ll k^+$ and $k^+$ scale as $\lambda^2$ or (3) $k$ has Glauber scaling. Assuming the gluon is emitted from a fast-moving nucleon in the $+z$ direction then (1) is highly improbable. Case (2) does not lead to any transverse effects like transverse gauge link or transverse broadening as all componets of $k$ scale similarly. Moreover soft gluons give rise to the  familiar Eikonal soft Wilson lines. In SCET soft contributions can be handled by field re-definitions \cite{Bauer:2001yt} and they get factorized from the collinear sector.
The remaining contribution comes from (3) where $p^-k^+\simeq \vert \vec{k}_\perp\vert^2\simeq k^2$.

In BJY, the exact scaling of the gluons is not specified. This issue will be addressed in some detail below.
For now, we continue the review of their work. By making use of the Chisholm's representation, one obtains the following form for the propagator, 
\bea
\frac{1}{2p^-k^++k_\perp^2-i\varepsilon}=i\int_0^\infty d\tau  e^{-i\tau(2p^-k^++k_\perp^2-i\varepsilon)},
\eea
where the left hand side is obtained from the limit $\tau=0$. With the above representation one is then able to carry out the integrations over $k^+$ for the amplitude of Fig.~\ref{FSI}. One then gets the gluon field $A^{\mu}(x)$ (in the mixed coordinate-momentum representation) evaluated at $A^{\mu}(x^-=2\tau p^-,x^+=0,\vec{k}_\perp)$. Now take the scaling limit $p^-\rightarrow \infty$ before performing the $\tau$ integrations. This sets the argument of the gluon field in the $-$ light-cone direction at infinity and all the remaining $\tau$-dependence is now in the exponent. Then perform the $\tau$ integration. This will result with a $\vert \vec{k}_\perp \vert^2$ in the denominator. By repeating the above set of manipulations one gets the following result for a multi-gluon exchange,
\bea
\label{bjj}
\langle p^-,N\vert j_{\nu}(0)\vert P \rangle &=&(-1)^n(i)^n(ig_s)^n{\bar u}(p^-){ \Pi}_{i=1}^n\langle N\vert \int \frac{d^2\vec{k}_i}{(2\pi)^2}\not\!A(\infty,\vec{k}_i)\frac{\sum_{j=1}^i\not\!{\vec k}_{j\perp}}{\vert \sum_{j=1}^i \vec{k}_{j\perp}\vert^2-i\varepsilon}
\gamma_\nu\psi(0)\vert P \rangle.
\eea

In light-cone gauge it is Eq.~\eqref{bjj} that gives the transverse gauge link. It is important to notice that the last result could be simply obtained by calculating, in full QCD, the amplitude for a Feynman diagram with arbitrary number of gluon attachments and then setting all the $k^+$ components of the gluon fields to $0$ wherever
they show up.  The last equation is only valid if one takes the scaling limit first which amounts to setting all the $k^+$ to $0$. This procedure clearly violates the Glauber scaling
as one has to maintain the relative scalings of $k^+$ and $\vert \vec{k}\vert^2/(2p^-)$ in tact.

The transverse gauge link that results from final-state interactions was also derived with somewhat different set of manipulations in \cite{stef} however the basic observation is still the same: it is gluons with vanishing $k^+$ that give rise to that gauge link. Another important issue related to the correct definition of the TMDPDF was raised in Refs.~\cite{stef,stef1}. There it was claimed that a soft factor, built up of soft Wilson lines, needs to be subtracted from the standard definition of the TMDPDF so as to get the desired features of the anomalous dimension of the TMDPDF. Soft factor subtractions were also discussed in the traditional literature of perturbative QCD (pQCD) \cite{ster2,col2,col3,col4} where this subtraction is aimed to avoid double-counting among mainly the collinear and soft contributions. In the effective field approach the double counting issue was treated within the ``zero-bin'' subtraction \cite{man} and later on a connection was made with the pQCD one \cite{idi1,idi2,lee}. It is interesting to see wether the arguments of soft subtraction based on the anomalous dimension arguments are equivalent to the double counting issue. We will not discuss this issue further here and we leave it to a future work.

\subsection{One Glauber Exchange}


We start our analysis of the TMDPDF in light-cone gauge ${\bar n}\cdot A=0$ by considering the relatively simple case of only one Glauber gluon interacting
with collinear quark. The kinematics are that of DIS:  the incoming quark is collinear in the $+z$ direction with $p^+$ 
large and of order of $Q$. This quark carries a longitudinal momentum fraction $x$ and carries transverse momentum $\kb$. The virtual photon moves in the $-z$ direction with momentum $q=(0,0,0,-Q)$. Thus after the hard interaction the quark has momentum with $p^-=Q/\sqrt{2}$ and transverse momentum $\kb$. The Glauber gluon has also $\kb$ so the outgoing quark has only $p^-$. 
Following the notation of BJY we consider the contribution of Fig.~\ref{FSI} to the matrix element
$\lc p^-, N \vert j_{\nu}(0)\vert P\rc$.

For Glauber gluons we now use the Feynman rules given in Fig.{~\ref{GLF1}}. We get
\bea
\label{one}
I_1&=& g_s~{\bar \xi}_{\bar n}\frac{\gamma^{\mu \perp}}{2p^-}\int \frac{dk^+}{2\pi}\int \frac{d^2\kb}{(2\pi)^2}\frac{\ks}{k^+ +\frac{\vert \kb \vert^2}{2p^-}-i\varepsilon} A_{\mu\perp}(k^+,\kb),
\eea
where we used ${\bar u}\frac{\not\!n \not\!{\bar n}}{2}={\bar u}$. We now invoke the Fourier transform
of $A(k^+,\kb)$:

\bea
\label{FT}
A_{\mu\perp}(k^+,\kb)&=&\int dx^-\int d^2\xb \tilde {A}_{\mu\perp}(x^-,\xb)e^{i(k^+x^--\kb\cdot\xb)}
\eea
and substitute for $A(k^+,\kb)$ into Eq.~(\ref{one}) and carry the $k^+$ integral by countour integration picking up the pole from $k^+=-\frac{\vert \kb \vert^2}{2p^-}+i\varepsilon$. The result is (from now on we drop the tilde on $\tilde A$),
\bea
\label{fin}
I_1&=&ig_s~{\bar \xi}_{\bar n}\frac{\gamma^{\mu \perp}}{2p^-}\int dx^-\theta(x^-)\int d^2\xb A_{\mu\perp}(x^-,\xb)\int \frac{d^2\kb}{(2\pi)^2}\ks e^{-i\left(\frac{\vert \kb \vert^2x^-}{2p^-}-\kb\cdot \xb\right)}.
\eea

In order to carry out the integral over $d^2\kb$ we complete the square in the exponent and shift the integration variable to $\vec k'_\perp=\kb+\frac{p^-}{x^-}\xb$. The resulting integral proportional to $\vec k'_\perp$ vanishes by symmetry. The remaining $d^2\kb$ integral is obtained as, 
\bea
&=&\int_{-\infty}^{\infty} \frac{d^2\kb}{(2\pi)^2}e^{-i\frac{\vert \kb \vert^2x^-}{2p^-}}=\frac{2p^-}{x^-}\frac{1}{(2\pi)^2}\left(\int_{-\infty}^{\infty} dk_xe^{\frac{-ik_x^2x^-}{2p^-}}\right)^2=\frac{2p^-}{x^-}\frac{1}{(2\pi)^2}\left((1-i)\sqrt{\frac{\pi}{2}}\right)^2\nonumber\\
&&=\frac{-i}{2\pi}\frac{p^-}{x^-}.
\eea
Introducing the above mentioned simplifications in Eq.~\eqref{fin}, the result for $I_1$ reads,
\bea
I_1&=&g_s\frac{1}{2\pi}\frac{\gamma^{\mu \perp}}{2p^-}\int dx^- \theta(x^-)\int d^2\xb \not\!\xb A_{\mu\perp}(x^-,\xb)\left(\frac{p^-}{x^-}\right)^2e^{\frac{-ip^{-} \xb \vert^2}{2x^-}}.
\eea
Let us consider the $x^-$ integral:
\bea
{\tilde I}=\int dx^- \theta(x^-) A_{\mu\perp}(x^-,\xb)\left(\frac{p^-}{x^-}\right)^2e^{\frac{-ip^{-} |\xb|^2}{2x^-}}
\eea
and perform the integration by parts. To do so we notice that  in light-cone gauge the $x^-$ dependence of $A_{\mu\perp}$ on $x^-$ is just a $\theta(x^-)$ \cite{jac1,jac2,jac3} (see Eq.~(\ref{glc}) below.) Moreover the contribution from a highly oscillating phase (obtained from the lower limit of the integral) would give a vanishing contribution upon integration over the transverse coordinates. The result of the integral over $x^-$ is then given as,
\bea
{\tilde I}=\frac{2ip^-}{\vert \xb \vert^2}\times A_{\mu\perp}(\infty,\xb).
\eea
Substituting the above result in the expression for $I_1$, we obtain, 
 \bea
I_1&=&ig_s\frac{1}{2\pi}\gamma^{\mu\perp}\int  d^2\xb  \frac{\not\!\xb}{\vert \xb \vert^2} A_{\mu\perp}(\infty,\xb).
\eea
In what follows, we decompose,  $\xb$ as $\xb=\vert \xb \vert \hat{n}_{\theta}$ where $\hat{n}_{\theta}\equiv (\cos \theta, \sin \theta)$. The integration measure is given as $d^2\xb= \vert \xb \vert d\vert \xb \vert d\theta $. Substitution in the expression for $I_1$ leads to the form, 
\bea
I_1&=&ig_s\frac{1}{2\pi}\gamma^{\mu\perp}\gamma^i \int_0^{2\pi} d\theta\int_0^{\infty} d\vert \xb \vert (\hat{n}_\theta)_iA_{\mu\perp}(\infty,\xb),
\eea
with $i=1,2$. Using the trivial relation: $\gamma^{\mu\perp}\gamma^i=\frac{1}{2}([\gamma^{\mu\perp},\gamma^i]+\{\gamma^{\mu\perp},\gamma^i\})$ and the fact that in light-cone gauge the gluon field  $A_{\mu\perp}$ at $x^- \ra \infty$ is a pure gauge~\cite{bjy}, we obtain, 
\bea
\label{glc}
A^{\perp}(x^- \ra \infty,\xb)&=&\theta(x^-)\vec{\nabla} \phi(r),
\eea
where $r\equiv \vert \vec{x}_\perp\vert$ and $\phi$ is an arbitrary scalar function. Then we have
\bea
A^{\perp}( x^- \ra \infty, \xb )=\frac{d\phi}{dr}\hat{n}_\theta, \label{dphi}
\eea
which shows that the gluon field $A_{\perp}$ is directed in 
the radial direction. With this, it is straightforward to show that the contribution from the commutator of the $\gamma$ matrices vanishes by 
symmetry. Thus we obtain, 
\bea
I_1&=&ig_s\frac{1}{2\pi}\int_0^{2\pi}d\theta\int_0^{\infty} d\vert \xb \vert (\hat{n}_\theta)_iA^{\perp}_i(\infty,\xb)
=ig_s\int_0^{\infty}d\vert \xb \vert \hat{n}\cdot A(\infty,\xb),
\eea
where, the second equality in the equation above is derived from the use of Eq.~\eqref{dphi}. 

The above analysis shows that it is indeed the Glauber gluons, arising from final state interactions, that build the transverse gauge link. In this sense, any (gauge invariant) effective field theory formulation, of the TMDPDF in particular (see e.g. Ref.~\cite{Chay:2007ty}) or semi-inclusive hadronic processes in general, requires the introduction of the Glauber mode in addition to the soft and collinear modes.

Two remarks are in order. We first notice that the above treatment could also be carried out with a multiple of Glauber gluons attachments. The power counting of such Feynman diagrams would still be a leading one since those contributions arise from the leading order Lagrangian. The sum of all those contributions would give the transverse gauge link. Secondly we notice that the power counting of the Glauber gluon field, in light-cone gauge could be read-off from Eq.~\eqref{dphi}. In Feynman gauge, the power counting of $A^+$ could also be read-of from explicit expressions (see, e.g., Eq.~(14) in \cite{ji1}.)


\section{conclusions}


Effective theories now constitute a mainstay in the collection of theoretical 
methods used to apply perturbative QCD to phenomenological questions. 
The Soft-Collinear-Effective-Theory has been identified as a rigorous and systematic 
effective approach in the application to phenomena involving hard jets in 
vacuum. In this article, we have instituted the first extension of this 
``leading twist'' effective theory to include power corrections from the 
medium. As a guide to understanding the effects of the medium on a jet,  
we have focused on a description of the rescattering encountered by a 
hard quark produced in deep-inelastic scattering on a nucleon in vacuum or 
within a large nucleus. While most of the results derived in this manuscript 
are immediately applicable to quark jets propagating through confined media, 
these may be straightforwardly extended to gluon jets as well as to propagation in deconfined media.

A jet in SCET is endowed with a very particular relation as regards the range of 
its different momentum components, 
\bea
p^\mu \equiv [p^+, p^-, \vp_\perp] \sim  Q [\lambda^2, 1 , \lambda],
\eea 
where, we have specified the case for a jet moving in the ($-$)-direction with $Q$, a hard scale and $\lambda$ a small parameter. 
The virtuality of this jet allows it to resolve modes in the medium with transverse momentum  
$k_\perp \sim \lambda Q$. If the forward or ($+$)-momentum components of these in-medium 
modes scale as $Q$ (or even as $\lambda Q$), this will result in an intermediate parton with large off-shellness 
of the order of $Q^2$ (or $\lambda Q^2$) and almost immediate hard radiation with large transverse momentum.
Such interactions will, no doubt, change the large momentum label of the propagating SCET mode and will 
be dealt with in more detail in a future effort. If the forward momenta scale as $\lambda^2 Q$, the off-shellness 
of the propagating mode remains within the scaling prescribed by SCET and as a result, the simplest 
extention to this vacuum theory is suggested in Sec.~II: the interaction between hard collinear quarks (or gluons) with 
gluons in the medium which scale as 
\bea
k^\mu  \sim  Q [\lambda^2, \lambda^2  , \lambda].
\eea 
Such gluons are referred to as Glauber gluons and in Sec.~III we have constructed the effective 
Lagrangian which describes their interactions with the hard collinear modes. Although we have 
denoted the scaling of the $k^-$ momentum to be $\lambda^2 Q$, it could indeed have any
scaling $k^- \lessapprox Q$ i.e., not be a hard collinear mode traveling in the ($-$)-direction. 
In the Breit frame with the medium moving with a large boost $\gamma \sim \lambda^{-2}$
in the ($+$)-direction, such modes are energetically disfavored.

Since Glauber modes carry a small fraction of the forward energy of the nucleon begin struck by the 
hard jet moving in the ($-$)-direction, they are quite pervasive and thus the inclusion of such modes 
and their interactions with the hard collinear modes is rather important. Such interactions 
occur continuously on a hard jet propagating through a dense medium. Given their off-shellness, 
SCET modes may traverse distances of the order of $(\lambda^2 Q)^{-1}$ before decay. When hard jets traverse 
large distances in dense matter, their total transverse momentum distribution is broadened. As a 
first application of the effective Glauber Lagrangian, this transverse broadening is derived for the 
case of DIS on a large nucleus in Sec.~IV. 
Multiple interactions 
with Glauber modes may eventually lead to the generation of off-shellness or transverse momenta 
beyond the range of applicability of SCET and a completely different effective theory will have
to be constructed. The transverse broadening as a function of the distance travelled, allows for an 
estimation of the range in size of media, within which, the effective theory will remain applicable.
This is estimated in Sec.~IV, with the aid of some phenomenological input.  It is argued that 
the derived effective theory has a wide range of applicability which may easily encompass jet propagation 
in the cold confined 
matter in large nuclei to that in hot deconfined matter created in high energy heavy-ion collisions.

In Sec.~V, as a second example of the role of Glauber gluons in hard processes, we have considered the treatment, 
in full QCD, of Belitsky, Ji and Yuan for the TMDPDF. In their calculation, 
they have shown that gluons with solely transverse momentum components build up a transverse gauge link which should 
be an integral part of the gauge invariant definition of the TMDPDF. Their analysis seemed to depend on taking the scaling 
limit first. In the current effort, we have demonstrated that the transverse gauge link may also be derived by keeping the 
Glauber scaling of gluon momenta between 
$k^+$ and $\vert \vec{k}_\perp \vert^2/2p^-$  explicit throughout the calculation. The full link structure, in any gauge, 
has been shown to arise from a combination of collinear and Glauber gluons. 

As a final remark we address certain situations where the  Glauber gluons do not contribute.
A standard example would be DIS on a nucleon with its related physical quantities: The quark 
form factor and the PDF. It has been demonstrated that the only relevant modes that produce 
the infra-red behavior of QCD for DIS are the soft and collinear (see, e.g, \cite{Sterman:1995fz} 
and references therein). Also for the factorization of the PDF itself (in the large $x$ limit) 
similar arguments and conclusions have been given in Ref.~\cite{JIMA}. The fact that Glauber gluons do not 
contribute to the factorization of DIS on a nucleon could, in principle, be shown 
(in the effective field theory approach) when one combines the soft, collinear
and Glauber in one framework and then certain cancellations of the Glauber contributions 
should become manifest. This is a somewhat more involved topic, which we leave for a future effort.

One may also compute the corrections to the single hadron inclusive cross section in DIS on a nucleon 
from the Glauber sector. In large nuclei, the produced jets tend to have a distribution in transverse 
momentum which is much wider than the case of DIS on a nucleon. 
It is well established that such corrections are not leading and are power suppressed, hence are unimportant in 
the case of DIS on a nucleon. 
This result is also consistent with the Glauber Lagrangian derived 
in the current manuscript. The magnitude of the correction from Glauber scattering may be estimated from 
Eq.~\eqref{A_scaling}, by setting $A = 1$. It is clear that the $\lc p_\perp^2 \rc$ generated is suppressed 
by $\lambda^2$ compared to that in a purely SCET process. Thus in the computation of the single hadron 
inclusive cross section from DIS on a nucleon, the contribution of the Glauber Lagrangian is power suppressed. 
In the case of the Drell-Yan process, the relevance of Glauber gluons is a more 
complicated issue. We will not discuss it further and refer the reader instead to Refs.~\cite{Collins:1983ju,Bodwin:1984hc,Doria:1980ak,Fries:2000da}.

In future efforts, the interaction between the Glauber modes emanating from the medium and the soft and collinear 
gluons of the SCET Lagrangian will have to be derived. This will represent the first complete theoretical 
description of jets with off-shellness in the range of $(\lambda Q)^2$ propagating through dense media. 
The setup of such a formalism will allow for the first systematic approach to such difficult problems such 
as factorization in hard jet production and modification in heavy-ion collisions. Embellishments of 
the Heavy-Quark-Effective-Theory (HQET) with Glauber modes will lead to more rigorous formulations of heavy-quark 
propagation in dense matter. Such extensions of SCET and HQET will lead to important advancements in 
our understanding of parton propagation and energy loss in dense matter and will no doubt play a leading role in 
the detailed theory and experimental comparison currently underway in DIS and heavy-ion collisions.


\section{Acknowledgments}


The authors wish to thank X.~Ji, C.~Kim, and X.~N.~Wang for helpful discussions. The authors also thank 
T.~Mehen and B.~M\"{u}ller for a careful reading of an earlier version of the manuscript and for discussions. 
The research of A.I. was supported in part by the U.S. Department of Energy
under grant numbers DE-FG02-05ER41368, DE-FG02-05ER41376, and DE-AC05-84ER40150.
The research of A.M. was supported in part by the U.S. Department of Energy
under grant numbers DE-FG02-05ER41367 and DE-FG02-01ER41190. 


\end{document}